\input BoxedEPS
\SetTexturesEPSFSpecial
\HideDisplacementBoxes

\documentclass[preprint,aps]{revtex4}

\newcommand{\BE}{\begin{equation}}
\newcommand{\EE}{\end{equation}}

\usepackage{graphicx}
\usepackage{dcolumn}
\usepackage{bm}
\def\bq{\begin{equation}}
\def\eq{\end{equation}}

\begin{document}
\draft
\title{Truncation effects in superdiffusive front propagation \\ with 
L\'evy flights}
\author{D. del-Castillo-Negrete
\thanks{e-mail: delcastillod@ornl.gov}}
\address{Oak Ridge National Laboratory \\ Oak Ridge TN, 37831-8071}
\pacs{82.40.Ck,02.50.Ey,05.40.Fb}

\begin{abstract}
A numerical and 
analytical study of the role of exponentially truncated L\'evy flights  in the 
superdiffusive propagation of fronts in reaction-diffusion systems is presented. The 
study is based on a variation of the Fisher-Kolmogorov equation where 
the diffusion operator is replaced by a $\lambda$-truncated fractional derivative of order 
$\alpha$ where $1/\lambda$ is the characteristic truncation length 
scale.  
For $\lambda=0$ there is no truncation and 
fronts exhibit exponential 
 acceleration and algebraic decaying tails. 
It is shown that for $\lambda \neq 0$ this phenomenology prevails 
in the intermediate asymptotic regime $\left( \chi t \right)^{1/\alpha} \ll x 
\ll 1/\lambda$ where $\chi$ is the diffusion constant. 
Outside the intermediate asymptotic regime, i.e. for $x > 1/\lambda$, 
the  tail of the front  exhibits the tempered decay 
$\phi \sim  e^{-\lambda x}/x^{\left( 1+\alpha\right)}\, $, the acceleration 
is transient, and the front velocity, $v_L$, approaches the terminal speed 
$v_* = (\gamma - \lambda^\alpha \chi)/\lambda$
as $t\rightarrow \infty$, where it is assumed that $\gamma > 
\lambda^\alpha \chi$ with $\gamma$ denoting the growth rate of the 
reaction kinetics.  However, the convergence of this process 
is algebraic, $v_L \sim v_* - \alpha 
/\left( \lambda t \right)$, which is very slow compared to the 
exponential convergence observed in the diffusive (Gaussian) case. 
An over-truncated regime in which the characteristic truncation 
length scale is shorter 
than the   length scale of the decay of the initial condition, $1/\nu$, is also identified. In this 
extreme regime, fronts exhibit exponential tails, $\phi \sim e^{-\nu x}$, and move at the 
constant velocity, $v=\left(\gamma - \lambda^\alpha \chi \right)/\nu$.

%
\end{abstract}
\maketitle

\section{Introduction}

Reaction-diffusion systems  have played a predominant role in the 
study of pattern formation and nonlinear dynamics in a large 
class of phenomena of interest to physics, biology, chemistry and 
engineering, see for example Refs.~\cite{cross,rd_biology} and 
references therein.  
One of the simplest  reaction-diffusion systems is the extensively 
studied Fisher-Kolmogorov model that describes the dynamics of a 
scalar field, $\phi$, in a one-dimensional domain,
\bq
\label{fk}
\partial_t \phi = \chi \partial_x^2 \phi + \gamma \phi \left ( 1 - 
\phi \right) \, ,
\eq
where $\chi$ denotes the diffusivity and $\gamma$ is a constant. 
The nontrivial dynamics of reaction diffusion systems  in general stems 
from the competition between the diffusivity and the non-linearity. In the 
case of the Fisher-Kolmogorov equation this competition leads to the 
propagation of fronts in which the stable, $\phi=1$, state advances 
through the destabilization of the $\phi=0$ unstable state.  

An important and often overlooked assumption in reaction-diffusion 
models is the use of Laplacian operators, $\chi \nabla^2$, for the description of 
transport. The use of these operators is motivated by the 
Fourier-Fick's model according to which the flux 
$q$  is assumed to be 
proportional to the gradient of the concentration, ${\bf q}=-\chi 
\nabla \phi$. This local prescription, together with the 
conservation law, $\partial_t \phi = -\nabla \cdot {\bf q}$, leads to the 
Laplacian, diffusive transport operator. From the statistical 
mechanics point of view this 
prescription is linked to the assumption that the underlying 
``microscopic'' transport process is driven by an uncorrelated, 
Markovian, Gaussian process.  However, despite its relative 
success,  experimental, numerical  
and theoretical evidence indicates that the diffusion model has limited 
applicability, see for example
Refs.~\cite{klafter_restaurant,anomalous_diffusion,book,metzler} and references therein. 
Therefore, a problem of considerable interest is the study of the 
role of anomalous diffusion, and L\'evy flights in particular, in reaction diffusion systems.

Early work on reaction-anomalous-diffusion systems  include 
Refs.~\cite{zanette,henry,vulpiani,del_castillo_2003}. Reference~\cite{zanette} studied 
bistable reaction processes and anomalous diffusion caused by L\'evy flights.
The interplay of  sub-diffusion and Turing instabilities was discussed in 
Ref.~\cite{henry}. The role of superdiffusive transport in the 
acceleration and algebraic decay of fronts was studied in the 
context  of a probabilistic approach in Ref.~\cite{vulpiani} and in the context of 
an equivalent fractional Fisher-Kolmogorov equation in 
Ref.~\cite{del_castillo_2003}. It is interesting to note that fronts in chaotic couple map lattices with 
long-range couplings exhibit an analogous phenomenology  as 
discussed in Ref.~\cite{torcini}. More recent works include:
the study of analytic solutions of fractional reaction-diffusion 
\cite{saxena_sols}; the study of a reaction-diffusion system with a bistable 
reaction term and directional anomalous diffusion \cite{barrio}; 
the study of the construction of 
reaction-subdiffusion equations \cite{sokolov_fronts};  the study of Turing 
instabilities \cite{henry_turing}; the study of the 
effect of superdiffusion on pattern formation selection in the Brusselator 
model \cite{sasha_turing}; and the study of the fractional  
Ginzburg-Landau and Kuramoto-Sivashinsky 
equations \cite{sasha_LG} among others.  

Evidence of L\'evy flights has been found in laboratory experiments, 
simple models and numerical studies of turbulent transport and the 
use of fractional diffusion models to describe these problems has 
been well-documented in the literature \cite{book}.
However, it is 
plausible that finite-size domain and decorrelation 
effects (among other effects) might have an impact  on the 
L\'evy flight dynamics. The evaluation of 
the  role of these ``non-ideal'' effects on reaction-anomalous-diffusion 
systems is a problem of considerable practical relevance. Of particular 
importance is to determine how, and to which degree, these 
effects might mask the underlying L\'evy  statistics. 
The effect of fluctuations caused by finite number of particles per 
volume on the superdiffusive propagation of fronts was studied in 
Ref.~\cite{brockman}. Here we focus on the role of truncation effects on
Levy flights driving superdiffusive front propagation. 

Asymptotic analysis plays an important role in the evaluation of non-ideal 
L\'evy flight effects. In particular, 
it is quite possible that because of  non-ideal 
effects  the statistics of the system will eventually converge to  Gaussian. 
However, the key issue is to determine the duration of the 
non-diffusive transient and the rate of convergence to Gaussian 
statistics. This  point is clearly illustrated 
in the ultraslow convergence to Gaussian statistics in the presence of 
truncated L\'evy flights. In this case
it has been  observed that although the statistics eventually converges to 
Gaussian (because of the central limit theorem)  a remarkably large 
number of steps is needed, and therefore the 
system effectively behaves non-diffusively in the intermediate asymptotic 
regime of practical interest
\cite{mantegna,kopone,shlesinger,feller,cartea_del_castillo_2007}. In the
present paper we explore to which degree a similar situation 
occurs in the case of front propagation. In particular, 
in Ref.~\cite{del_castillo_2003} it was shown that in the 
 fractional Fisher-Kolmogorov equation fronts decay algebraically and 
exhibit exponential acceleration. The goal of this paper is to 
present a numerical and analytical asymptotic study of the effect of truncation 
on these phenomena. One problem of special interest is to determine 
if there is an intermediate asymptotic regime where the effects of 
truncation are negligible and where the fronts accelerate and exhibit algebraic 
tails.
Outside such intermediate asymptotic regime it is expected that the truncation effects 
will become dominant and that as $t \rightarrow \infty$ the front  
dynamics will eventually approach in some sense the diffusive Fisher-Kolmogorov 
dynamics. However, the key issue is how long will this take. This  
bring us to the second problem of interest in this paper which to determine the rate of convergence to 
the  constant velocity and exponential tails characteristic of the 
diffusive front propagation regime. 

Our approach is based on the use of truncated fractional diffusion 
operators.  Fractional derivatives  provide a powerful framework 
to model non-diffusive transport processes
~\cite{metzler,book}. These operators incorporate long-range, non-local 
transport through the use of slowly decaying kernels. In 
particular, in 
fractional diffusion the Laplacian is replaced by an integral 
operator of the form $\partial^2_x\int \phi(x',t) K(x-x') dx'$, where the 
kernel $K$ has the asymptotic behavior, $K\sim 
x^{-\alpha+1}$, for $1<\alpha<2$.
In the context of the Continuous Time Random Walk model the 
exponent $\alpha$ is related to the stability index of the underlying 
L\'evy process \cite{metzler}. However, a potential drawback of this 
description is that $\alpha$-stable L\'evy processes 
have divergent second moments because their corresponding densities 
decay as $\sim x^{-(1+\alpha)}$. This issue has motivated the 
consideration of truncated L\'evy processes  that exhibit long 
range dynamics  while preserving the finiteness of some \cite{sokolov}  or all  
moments \cite{mantegna,kopone,cartea_del_castillo_2007}. In the case of exponentially truncated 
processes, the L\'evy density decays as $\sim x^{-(1+\alpha)} 
e^{-\lambda x}$, and for  $\lambda \neq 0$ all the moments are 
finite. In the present paper we limit attention to this type of 
processes and study the truncated fractional Fisher-Kolmogorov in 
which the Laplacian operator in Eq.~(\ref{fk}) is replaced by the 
truncated fractional diffusion operator proposed in 
Ref.~\cite{cartea_del_castillo_2007}.
 
The organization of the rest of the paper is as follows. The next 
section reviews material on truncated L\'evy flights, defines the 
$\lambda$-truncated fractional derivatives and discusses the fundamental
solutions and scaling properties of the truncated fractional diffusion 
equation. Section~III contains the core of the numerical results obtained from 
the integration of the truncated fractional Fisher-Kolmogorov 
equation for front-type initial conditions. Section~IV presents an analytical asymptotic study 
based on the leading 
edge approximation. The conclusions are presented in 
Sec.~V. 

\section{Fractional diffusion with truncated  L\'evy flights}
In the standard fractional diffusion model the transport  of a scalar 
$\phi$ is governed by the equation
\begin{equation}
\label{eq_23}
_{0}^{c}D_t^\beta \phi(x,t)= -a \partial_x \phi +
c \left[ l \,
_{-\infty}D_x^\alpha +  r \, _{x}D_{\infty}^\alpha \right] \, \phi \, ,
\end{equation}
where  we have included in addition to the spatial and temporal 
fractional operator an advective term.
The operator on the left-hand side is the regularized (in the Caputo 
sense \cite{podlu_1999}) fractional time derivative
\begin{equation}
\label{eq_11}
_{0}^{c}D_t^\beta \phi= \frac{1}{\Gamma(1-\beta)} \, \int_0^t
\frac{\partial_\tau \phi}{(t-\tau)^\beta}\,d\tau \, ,
\end{equation}
with $0<\beta<1$ and the operators on the right hand side of Eq.(\ref{eq_23}) are the left and right 
Riemann-Liouville fractional derivatives
\cite{podlu_1999,samko_1993}
\bq
_{a}D_x^\alpha \phi= \frac{1}{\Gamma(m-\alpha)}\,
\frac{\partial^m}{\partial x^m}\,
\int_a^x\, \frac{\phi}{(x-y)^{\alpha+1-m}}\, dy \, ,
\eq
\bq
_{x}D_b^\alpha \phi = \frac{(-1)^m}{\Gamma(m-\alpha)}\,
\frac{\partial^m}{\partial x^m}\,
\int_x^b\, \frac{\phi}{(y-x)^{\alpha+1-m}}\, dy \, ,
\eq
with $m-1 \leq \alpha < m$.
The weighting factors $l$ and $r$ are defined as
\begin{equation}
\label{eq_24}
l=-\frac{(1-\theta)}{2 \cos(\alpha \pi/2)}\, , \qquad
r=-\frac{(1+\theta)}{2 \cos(\alpha \pi/2)} \, ,
\end{equation}
with $-1<\theta<1$.
According to Eqs.~(\ref{eq_23}) and (\ref{eq_24}), the parameter $\theta$ determines the 
degree of asymmetry of the fractional operator. For $\theta=0$ the 
contributions of left and right derivatives are equal and the 
operator is symmetric. In the extremal, fully asymmetric case of 
main interest in this paper, $\theta=-1$, and only the left derivative 
is present in the diffusion operator. 
Equation~(\ref{eq_23}) describes the fluid limit of a
continuum time random walk (CTRW) in the case when the waiting time 
distribution function exhibits algebraic decay of the form, $\psi 
\sim t^{-1-\beta}$, and the particle jumps follow an $\alpha$-stable 
L\'evy distribution, see for example \cite{metzler} and references therein.

The fractional equation (\ref{eq_23}) has found applications in 
several areas of 
physics, engineering and biology. For a recent discussion on
the basic theory and applications of fractional diffusion 
see \cite{book}. 
However more general 
transport equations can be obtained when a wider class of stochastic 
processes are considered. In particular, in the case when the 
particle jump probability density function corresponds to an 
exponentially truncated 
(tempered) distribution with characteristic exponent of the form
\cite{kopone}
\begin{eqnarray}
\label{eq_26}
\Lambda_{ET}=i a k -
\frac{c}{2 \cos(\alpha\pi/2)} \left\{
\begin{array}{ll}
(1+\theta) (\lambda +ik )^{\alpha }+
  (1-\theta) (\lambda -ik)^{\alpha }
- 2\lambda ^{\alpha }
\mbox{,} \\
(1+\theta)(\lambda +ik )^{\alpha }+
(1-\theta)(\lambda -ik )^{\alpha }-
2 \lambda^{\alpha }- 2 i k\alpha \theta \lambda ^{\alpha -1}  \, ,
\end{array}
\right.
\end{eqnarray}
for $0<\alpha<1$ and $1<\alpha\leq2$ respectively, the fluid limit of 
the corresponding CTRW leads to the equation~\cite{cartea_del_castillo_2007}
\begin{equation}
\label{eq_27}
_0^cD_t^\beta \phi = -V \partial_x \phi +
c {\cal D}_x^{\alpha,\lambda} \phi -\mu \phi   \,  , 
\end{equation}
where the $\lambda$-truncated fractional derivative operator of order $\alpha$,
${\cal D}_x^{\alpha,\lambda}$,
is defined as
\begin{equation}
\label{eq_28}
{\cal D}_x^{\alpha,\lambda} = l e^{-\lambda x}\,
_{-\infty}D_x^\alpha\, e^{\lambda x}\, +r e^{\lambda
x}\, _{x}D_{\infty}^\alpha\, e^{-\lambda x}  \, .
\end{equation}
The effective advection velocity is  $V=a$ for $0<\alpha<1$, and $V=a-v$ for  $1<\alpha<2$ with 
\bq
\label{eq_v} 
v=  \frac{c \alpha \theta \lambda^{\alpha-1}}{\left|
\cos\left( \alpha \pi /2\right)\right|} \, , \end{equation} and
\begin{equation}
\label{eq_29}
  \mu = -\frac{c \lambda^\alpha}{\cos\left(
\alpha \pi /2\right)}\, .
\eq
According to Eq.~(\ref{eq_v}), in the case $1<\alpha<2$, the 
truncation gives rise to a drift that depends on the asymmetry of 
the process. 
The parameter $\lambda$ determines the truncation of the tempered 
L\'evy process whose corresponding L\'evy density is given by \cite{kopone} 
\begin{eqnarray}
\label{eq_25}
w_{ET}(x)=\left\{
\begin{array}{ll}
c\frac{(1+\theta)}{2}\left| x\right| ^{-(1+\alpha) }e^{-\lambda |x|}
& \mbox{ for
}x<0\mbox{,}
\\ c\frac{(1-\theta)}{2}x^{-(1+\alpha) }e^{-\lambda x} & \mbox{for }x>0,
\end{array}
\right.
\end{eqnarray}
  $0<\alpha\leq 2$, $c>0$,
$-1\leq\theta\leq 1$ and $\lambda\geq 0$. 
As expected, for $\lambda=0$, Eq.~(\ref{eq_25}) reduces to the 
$\alpha$-stable density, and  Eq.~(\ref{eq_27}) reduces to Eq.~(\ref{eq_23}).

The general solution of Eq.~(\ref{eq_27}) for an initial condition 
$\phi(x,t=0)=\phi(x,0)$ is
\bq
\label{sol}
\phi(x,t)= \int_{-\infty}^{\infty} G_{\alpha, \beta, \theta, 
\lambda}(x-x',t) \phi(x',0) dx' \, ,
\eq
where $G_{\alpha, \beta, \theta, \lambda}$ is the Green's function or propagator which 
corresponds to the solution with initial 
condition $\phi(x,t=0)=\delta(x)$. The 
sub indices of $G$ state explicitly that in general the solution depends 
on four parameters: the order of the fractional derivative in space 
$\alpha$, the order of the fractional time derivative $\beta$, the 
asymmetry of the fractional operator $\theta$, and the truncation 
$\lambda$. 
Using the Fourier transform properties of the truncated 
fractional derivative it follows that
\begin{equation}
\label{eq_17}
G_{\alpha, \beta, \theta, \lambda}(x,t)=\frac{1}{2\pi} \int_{-\infty}^\infty e^{-ik
x}E_{\beta}\left[t^{\beta} \Lambda_{ET} (k)\right]dk \, ,
\end{equation}
where $E_\beta$ denotes the Mittag-Leffler function and 
$\Lambda_{ET}$ is given in Eq.~(\ref{eq_26}). At first sight the linear term, $-\nu \phi$, on the right 
hand side of Eq.~(\ref{eq_27}) seems strange and likely to give rise to a 
unphysical damping of the transported field, $\phi$. However, quite 
to the contrary, 
this term is critical to guarantee the conservation of $\phi$. When $\phi$ is interpreted as a probability 
density function, this term guarantees the normalization and 
conservation of the total probability. One way to see this is to note 
that this term comes from the term proportional to $-2\lambda^\alpha$ 
in (\ref{eq_26})  that implies $\Lambda_{ET}(k=0)=0$. 
According to Eqs.~(\ref{sol})  and (\ref{eq_17})  
$\int \phi(x,t) dx = E_\beta\left[t^\beta \Lambda_{ET}(k=0)\right] 
\int \phi_0(x) dx$ which guarantees the conservation of $\phi$ 
provide $\Lambda_{ET}(k=0)=0$, since $E_\beta(0)=1$.

In this paper we focus on the special case of Eq.~(\ref{eq_27}) with 
$1<\alpha<2$, $\beta=1$, and $\theta=-1$. Also, we assume an 
advection velocity $a=v$, which  results in the following asymmetric, truncated 
fractional equation 
\bq
\label{trun_asy}
\partial_t \phi = \chi \left[ e^{-\lambda x}\,
_{-\infty}D_x^\alpha\, \left( e^{\lambda x} \phi \right) - \lambda^{\alpha}\phi \right] 
\, ,
\eq
with $\chi=c/\left| \cos\left(\alpha \pi/2\right) \right |$. We 
restrict attention to this special case because it corresponds 
to the truncated version of the $\alpha$-stable asymmetric fractional 
operator used in the front acceleration problem discussed in 
Ref.~\cite{del_castillo_2003}. For this case, the general  solution 
in Eq.~(\ref{eq_17}) reduces to 
\bq
\label{trunc_green}
G_{\alpha,1,-1,\lambda}= \frac{1}{2 \pi} \int_{-\infty}^{\infty} \, e^{-ikx + \chi t \left[ 
\left( \lambda - i k\right)^{\alpha} - \lambda^{\alpha}
\right]}\, dk \, ,
\eq
which can be equivalently written as
\bq
G_{\alpha,1,-1,\lambda}= e^{-\lambda x-\chi \lambda^\alpha t} 
\left( \chi t \right)^{-1/\alpha} \hat{G}_{\alpha,1,-1,0}(\eta) \, ,
\eq
where
\bq
\label{green_lambda_0}
\hat{G}_{\alpha,1,-1,0}(\eta)=\frac{1}{2\pi} \int_{-\infty}^{\infty} 
e^{i^\alpha k^\alpha + i k \eta} \, dk \, ,
\eq
is the Green's function of the asymmetric, 
$\alpha$-stable ($\lambda=0$) fractional diffusion  equation, in 
terms of the similarity variable $\eta=  x \left( \chi t \right)^{-1/\alpha}$.
From here, using the asymptotic expression, 
$\hat{G}_{\alpha,1,-1,0}(\eta) \sim \eta^{-1-\alpha}$  for $\eta>0$
\cite{taqu,mainardi_2001} it follows that
\bq
\label{trunc_green_as}
G_{\alpha,1,-1,\lambda} \sim \chi t  e^{-\chi \lambda^\alpha t} 
\,\frac{e^{-\lambda x}}{x^{1+\alpha}} \, ,
\eq
for $x>0$ and $x \gg \left( \chi t \right)^{1/\alpha}$. 
For the decay of the left tail,  we use the asymptotic expansion
$\hat{G}_{\alpha,1,-1,0}(\eta) \sim \left | \eta \right |^{a_2} \, e^{-b_2 
\left | \eta \right |^{c_2}}$
for $\eta<0$ and $|\eta| \gg 1$, where $a_2=(2-\alpha)/(2 (\alpha-1))$, 
$b_2=(\alpha-1)\alpha^{\alpha/(\alpha-1)}$ and 
$c_2=\alpha/(\alpha-1)$  \cite{mainardi_2001}, to conclude
\bq
\label{green_1}
G_{\alpha,1,-1,\lambda} \sim \left(\chi t  \right)^{-(a_2+1)/\alpha}
e^{-\chi \lambda^\alpha t} \left | x \right |^{a_2}\, e^{-b_2 
\left(\chi t \right)^{-c_2/\alpha}\, |x|^{c_2}+ \lambda |x|} \, ,
\eq
for $x<0$ and $|x| \gg \left( \chi t \right)^{1/\alpha}$.  Since we 
are assuming that $1<\alpha <2$, it follows that $c_2>1$ and the $-|x|^{c_2}$ term in 
the exponent dominates the $\lambda |x|$ term, leading to a faster than 
exponential decay of the left tail for any value of $\lambda$. 
Figure~1 shows plots of the Green's 
function in Eq.~(\ref{trunc_green}) for $\alpha=1.5$ and different 
values of $\lambda$, along with the asymptotic approximation in 
Eq.~(\ref{trunc_green_as}). 

An important property of truncated L\'evy flights, 
originally discussed in Refs.~\cite{feller,mantegna,shlesinger}, is the ultraslow 
convergence to Gaussian statistics.  
According to this result, the  crossover time for Gaussian 
behavior to appear, $\tau_c$, scales as
\bq
\label{tauc}
\tau_c \sim \chi^{-1} \lambda^{-\alpha}\, ,
\eq
as expected, as $\lambda \rightarrow 0$, $\tau \rightarrow \infty$. 
When  memory effects are incorporated using fractional time 
derivatives, the crossover dynamics is richer. In particular, for $2 
\beta /\alpha >1$, $\tau_c \sim 
\chi^{-1/\beta}\lambda^{-\alpha/\beta}$  signals the 
crossover from superdiffusive to subdiffusive dynamics \cite{cartea_del_castillo_2007}.
The time scale $\tau_c$ will play an important role in the dynamics of the 
fronts. 

\section{Front propagation in the presence of truncated L\'evy 
flights: numerical results}
To study the role of truncation in the superdiffusive  acceleration 
of fronts due to L\'evy flights we consider 
the fractional Fisher-Kolmogorov equation originally introduced in 
Ref.~\cite{del_castillo_2003} and substitute the Riemann-Liouville 
fractional derivatives (which correspond to $\alpha$-stable L\'evy 
processes) by the truncated fractional derivative in 
Eq.~(\ref{eq_28}).
In the most general case the 
resulting equation is
\begin{equation}
\label{tffk_eq}
_0^cD_t^\beta \phi = -V \partial_x \phi +
c {\cal D}_x^{\alpha,\lambda} \phi -\mu \phi  +\gamma \phi \left( 1- 
\phi \right) \,  .
\end{equation}
However, as mentioned before, to compare the results with those in 
Ref.~\cite{del_castillo_2003}  we will  restrict attention to asymmetric, truncated 
fractional diffusion operators of the form in Eq.~(\ref{trun_asy}) and consider
\bq
\label{tffk}
\partial_t \phi = \chi \left[ e^{-\lambda x}\,
_{-\infty}D_x^\alpha\, \left( e^{\lambda x} \phi \right) - \lambda^{\alpha}\phi \right] 
+ \gamma \phi \left( 1 - \phi \right )
\, .
\eq
In this section we present results obtained from the numerical 
integration of Eq.~(\ref{tffk}).   We assume $\phi =A$,  for $x<0$, where $A$ is a 
constant  and  discretize the fractional derivative in the $x \in 
(0,1)$ domain using  the Grunwald-Letnikov representation.
Details of  finite-difference  methods for the solution of fractional diffusion 
equations can be found in \cite{del_castillo_2006,lynch}. 
In all the numerical simulations we 
consider $\alpha=1.5$, $\theta=-1$, $\gamma=1$, 
$\chi=5\times 10^{-7}$, and initial conditions of 
the form 
\bq
\label{ic}
\phi(x,t=0)=e^{-\nu x}\, ,
\eq
where $\nu$ is a constant.   

Figure~\ref{fig_fronts}-(a) shows snapshots of the front profile 
$\phi$ as function of $x$ at different times in the $\alpha$-stable 
($\lambda=0$) case obtained from the numerical solution of Eq.~(\ref{tffk}). In this case the front  
exhibits an algebraically decaying tail 
which in log-log scale manifests as a straight line \cite{del_castillo_2003}. 
Figure~\ref{fig_fronts}-(b) shows that the algebraic decay of the 
tail remains for small values of $\lambda$. In fact, as we will 
discuss in the next section, there is an intermediate asymptotic 
regime in which the role of truncation is negligible. 
Outside the intermediate asymptotic 
regime the  effect of 
the truncation depends critically on the ratio of the length scale of 
the truncation, $1/\lambda$, and  length scale, $1/\nu$, of the initial condition. 
When, $\lambda 
< \nu$, i.e., when the initial condition decays faster than the 
truncation, the tail of the front scales as $\phi \sim 
x^{-1-\alpha}\,e^{-\lambda x}$ as shown in Fig.~\ref{fig_fronts}-(c). 
On the other hand, when  $\lambda > \nu$, the truncation effects 
dominate and L\'evy statistics seems to 
have no effect on the dynamics of the front which, as shown in 
Fig.~\ref{fig_fronts}-(d) (note the log-normal scale),
exhibits the usual exponential decay of the diffusive Fisher-Kolmogorov model. 

When the front exhibits ``rigid'' propagation with a constant 
velocity, as  in the  diffusive 
Fisher-Kolmogorov case, it is straightforward to define and 
numerically compute the front speed. However, when the front 
accelerates and deforms, as it is the case in the $\alpha$-stable and 
truncated fractional Fisher-Kolmogorov 
equation, computing the front speed is not straightforward. One way 
to approach this problem is to consider the Lagrangian trajectory of 
the front's tail. For a given value of 
$\phi_0$, we define the Lagrangian trajectory, $x_L=x_L(t)$, of the 
front according to $\phi(x_L(t),t)=\phi_0$. Given $x_L$ we define the Lagrangian velocity as
$v_L=d x_L/dt$ and the Lagrangian acceleration as
$a_L=d v_L/dt$.  

Figure~\ref{fig_x_l} 
shows space-time plots of the Lagrangian trajectories of fronts 
corresponding to $\phi_0=10^{-6}$. Due to the relatively small value 
of $\phi_0$, these orbits follow the Lagrangian dynamics of the 
fronts' leading edge. The solid lines denote the numerical values for 
different values of $\lambda$ and the dashed lines denote the result of the 
asymptotic analytic calculation that will be discussed in the next 
section. For $0\leq \lambda <1$ it is observed that the front 
propagates very fast. However,  as $\lambda$ increases the speed of the 
front is  reduced. The corresponding Lagrangian 
velocities and accelerations are shown in Figs.~\ref{fig_v_l} and 
\ref{fig_a_l}. For small 
$\lambda$, the font acceleration grows monotonically. However, 
for larger values of $\lambda$ the 
Lagrangian velocity approaches a terminal velocity as $t 
\rightarrow \infty$. Note that as $\lambda$ increases, the 
terminal velocity is smaller and the convergence is faster. 

As shown 
in Fig.~\ref{fig_a_l}, in all cases the acceleration exhibits a pulse like behavior in the time evolution.
The time of peaking of the pulse increases (approximately exponentially) with $\lambda$ and the 
amplitude of the acceleration's peak decreases (approximately exponentially) with $\lambda$.
However, the key feature to note  is the evolution of the acceleration following 
the transient short pulse. For small $\lambda$ the front 
acceleration exhibits a monotonic increase whereas for larger values  
the acceleration exhibits a transient growth
followed by an eventual 
decay.  Figure~\ref{fig_v_l_a_l_sca} shows the asymptotic scaling behavior of the 
Lagrangian velocity and acceleration. The solid line curves denote 
the numerical results for different values of $\lambda$. It is  
observed that the convergence to the terminal velocity 
scales as $v_*-v_L \sim 1/t$ and the decay of the transient 
acceleration scales as $a _L\sim 1/t^2$. These numerical results are in agreement with the 
asymptotic scaling, shown with dashed line curves, that will be 
discussed in the following section.

\section{Leading edge asymptotic dynamics: analytical results}

\label{sec_leading_edge}

In this section we present analytical results describing the 
asymptotic behavior of fronts in the truncated fractional 
Fisher-Kolmogorov equation. The results are based on the 
leading edge approximation. This approximation exploits the idea that at the 
leading edge of the front  $\phi \ll 1$, and therefore in this 
region the nonlinear reaction term can be linearized around 
$\phi=0$, resulting in the linear equation
\bq
\label{tffk_lin_eq}
\partial_t \phi = \chi e^{-\lambda x}\,
_{-\infty}D_x^\alpha\, \left( e^{\lambda x} \phi \right) 
+\left(\gamma - \chi \lambda^{\alpha}\right) \phi 
\, .
\eq
Note that the truncated fractional derivative has a direct effect on 
the growth rate through the term $-\chi \lambda^\alpha$.  As 
discussed in the previous section, this term is
key to guarantee the conservation of the transported field. 
Two characteristic time scales can be distinguished in this problem: 
the cross-over to Gaussian statistics time scale, $\tau_c \sim 1/\left(\chi 
\lambda^\alpha \right)$, and the reaction time scale, $\tau_r=1/\gamma$. 
We will assume that $\tau_c >\tau_r$ to guarantee that the 
effective reaction constant, $\gamma_{eff}=\gamma - \chi \lambda^\alpha$,  
is positive as needed for 
the excitation and propagation of ``pull'' type fronts.

Substituting 
\bq
\label{phi_psi}
\phi = e^{-\lambda x +\left( \gamma - \chi \lambda^\alpha \right) t}\psi(x,t) \, ,
\eq
into Eq.~(\ref{tffk_lin_eq})  gives the 
asymmetric, $\alpha$-stable fractional diffusion equation
\bq
\partial_t \psi= \chi 
_{-\infty}D_x^\alpha \psi
\, . 
\eq
The general solution of this equation for an 
initial condition $\psi(x,t=0)=\psi_0(x)$ can be written as
\bq
\label{green_psi}
\psi(x,t) = \int_{-\infty}^{\infty} \hat{G}_{\alpha,1,-1,0}(\eta) \, \psi_0 \left[ x - 
\left(\chi t \right)^{1/\alpha} \eta \right ] d \eta \, ,
\eq
where $\hat{G}_{\alpha,1,-1,0}$ given in Eq.~(\ref{green_lambda_0}).

Consistent with the numerical simulations, we 
consider an initial condition of the form
$\phi(x,t=0)=A$ for $x <0$ and $\phi(x,t=0)=e^{-\nu x}$, where $A$ 
and $\nu$ are constants. Substituting the corresponding initial 
condition for $\psi$, according to Eq.(\ref{phi_psi}), into the 
solution in Eq.(\ref{green_psi}) 
we get
\bq
\psi = e^{-\left(\nu - \lambda \right)x}\, \int_{-\infty}^{x/\tau} \, 
e^{\left(\nu - \lambda \right) \tau \eta} \hat{G}_{\alpha,1,-1,0}(\eta) d \eta + 
A e^{\lambda x}\, \int_{x/\tau}^{\infty} \hat{G}_{\alpha,1,-1,0}(\eta) e^{-\lambda \tau 
\eta} d \eta \, ,
\eq
where we have defined $\tau = \left(\chi t\right)^{1/\alpha}$.
In terms of $\phi$ the solution can be written as
\bq
\label{lead_edge}
\phi= e^{-\nu x+ \left(\gamma - \chi \lambda^\alpha \right) t}\, 
{\cal  I}_1 + A e^{\left(\gamma - \chi \lambda^\alpha \right) 
t}{\cal  I}_2 \, , 
\eq
where
\bq
\label{integrals}
{\cal  I}_1=\int_{-\infty}^{x/\tau} \, 
e^{\left(\nu-\lambda \right) \tau \eta} \hat{G}_{\alpha,1,-1,0}(\eta) d \eta \, , \qquad
{\cal  I}_2=\int_{x/\tau}^{\infty} \hat{G}_{\alpha,1,-1,0}(\eta) e^{-\lambda \tau 
\eta} d \eta \, .
\eq
The goal of this section is to study the 
asymptotic behavior of ${\cal I}_1$ and ${\cal I}_2$ for $x/\tau 
\rightarrow \infty$. 
Before getting in the calculation of main interest here, it is 
instructive to consider first the diffusive (Gaussian) and fractional 
($\alpha$-stable) limits  of the leading edge solution in Eq.(\ref{lead_edge}).

\subsection{Diffusive case}
In the diffusive (Gaussian) case, $(\alpha, \theta, 
\lambda)=(2,0,0)$, the leading edge solution in Eq.~(\ref{lead_edge}) reduces 
to 
\bq
\phi = e^{-\nu x + \gamma t}
\int_{-\infty}^{x/\sqrt{\chi t}} \hat{G}_{2,1,0,0}(\eta) e^{\nu 
\,\eta \, \sqrt{\chi t}} 
d \eta +
A\, e^{\gamma t} \, 
\int_{x/\sqrt{\chi t}}^{\infty} \hat{G}_{2,1,0,0}\, d \eta \, ,
\eq
where $\hat{G}_{2,1,0,0}=1/(2 \sqrt{\pi}) e^{-\eta^2/4}$ is the 
Gaussian propagator. Introducing the normal probability distribution 
function $P(z)=(1/\sqrt{2 \pi}) \int_{-\infty}^{z} e^{-u^2/2} d u$,
\bq
\phi = e^{-\nu x + \left(\gamma +\nu^2 \chi \right) t}
P\left( \frac{x-2 \nu \chi t}{\sqrt{2 \chi t}}\right) +
A e^{\gamma t} \left[1 -P\left( \frac{x}{\sqrt{2 \chi t}}\right) 
\right] \, .
\eq
Using the asymptotic expansion $P(z) \sim 1-(1/\sqrt{2 \pi}) 
e^{-z^2/2} /z$, we conclude that in the limit 
$\eta=x/\tau \gg 1$, 
$x > 2 \nu \chi t$, 
\bq
\label{gauss}
\phi \sim e^{-\nu \left( x - c t \right)} + \sqrt{\frac{\chi t}{\pi}} \left[ 
\frac{A}{x}-\frac{1}{x-2\nu \chi t} \right] 
e^{-\frac{\gamma}{c_m^2 t} \left( x - c_m  t\right)\left( x + c_m 
t\right)} \, ,
\eq
where
\bq
\label{diff_speed}
c=\frac{\gamma}{\nu} + \nu \chi \, .
\eq
That is, in this case, the leading edge exhibits the well-known asymptotic
exponential dependence, $\phi \sim e^{-\nu \left( x - c t \right)}$, 
and the front propagates at the constant speed $c$ with  $c_m=2 \sqrt{\gamma 
\chi}$ corresponding  to the minimum  speed 
achieved for $\nu=\sqrt{\gamma/\chi}$. Note that according to 
Eq.~(\ref{gauss}) in this case the convergence to constant speed  is exponentially fast, 
i.e., the second term in the asymptotic expansion scales as $\sim 
e^{-a \left( x/\tau \right) ^2}$. This result will be contrasted 
below with the much slower convergence in the case of truncated L\'evy 
flights. 

\subsection{Fractional case}
The fractional ($\alpha$-stable) case was discussed in 
Ref.~\cite{del_castillo_2003}. In this case,  the leading edge solution is given by 
Eqs.~(\ref{lead_edge}) and (\ref{integrals}) with $\lambda=0$ and the 
corresponding leading asymptotic behavior is
\bq
\label{alpha-stable}
\phi \sim\chi t e^{\gamma t} \left[ \frac{A}{\alpha} x^{-\alpha} + 
\frac{1}{\nu} x^{-1-\alpha} + \dots \right] 
\, ,
\eq
where as mentioned before the constant $A$ relates to the boundary 
condition $\phi=A$ for $x <0$. 
The critical difference with the Gaussian case is the algebraic 
decay of the leading edge accompanied by the exponential acceleration 
of the front.   Note that when $A \neq 0$, (which was the 
case considered in Ref.~\cite{del_castillo_2003}) the front tail  exhibits the 
decay $\phi \sim 1/x^{\alpha}$. However, when $A=0$, the front decays 
faster, $\phi \sim 1/x^{\alpha+1}$. Figure~\ref{fig_fronts}-(a) shows a 
numerical verification of this scaling. This result will be contrasted 
with the truncate L\'evy flights case, where as in the Gaussian case, 
it will be shown that the rate of decay of the front's tail is 
independent of $A$.  

In the Gaussian case,  the 
spatio temporal evolution of the leading edge depends  to leading 
order on the variable $x-ct$ which implies a ``rigid'' translation of 
the exponential tail of the front and allows the interpretation of $c$ 
as the front. However, in the non-Gaussian case each point of the leading edge 
moves at a different speed and the tail does not translate 
rigidly. As discussed in the previous section we circumvent this 
problem by 
considering the Lagrangian trajectory 
$x_L=x_L(t;\phi_0)$ of a point in the leading 
edge of the front such that $\phi(x_L(t),t)=\phi_0$ where $\phi_0 
\ll 1$. 
According to Eq.~(\ref{alpha-stable}), in the $\alpha$-stable case, 
for $A=0$,
\bq
\label{asy_x}
x_L(t)=C \exp\left[{\frac{1}{1+\alpha} \left ( \gamma t + \ln t 
\right )} \right] \, ,
\eq
where $C$ is a constant that depends on $\phi_0$, $\chi$ and $\nu$, 
and 
\bq
\label{asy_v}
v_L(t)=\frac{\gamma}{1+\alpha} \left( \frac{\chi t}{\phi_0 
\nu}\right)^{\frac{1}{1+\alpha}}\left[ \frac{1}{\gamma t} +1 \right ] 
e^{\left(\frac{\gamma}{1+\alpha}\right) t} \, ,
\eq
which implies an unbounded, exponential growth of the front speed. 
For large $t$, the corresponding leading-order behavior of the  front acceleration is
\bq
\label{asy_a}
a_L(t)=\left( \frac{\gamma}{1+\alpha} \right)^2 \left( \frac{\chi 
t}{\phi_0 \nu}\right)^{\frac{1}{1+\alpha}} e^{\left( 
\frac{\gamma}{1+\alpha}\right) t} \, .
\eq
As shown in Figs.~\ref{fig_x_l}, \ref{fig_v_l}, and \ref{fig_a_l} the asymptotic 
results in Eqs.~(\ref{asy_x})-(\ref{asy_a}) are in good 
agreement with the numerical results discussed in the previous 
section. 

\subsection{Truncated case}

Going back to the general truncated fractional case, we consider  
first the asymptotic behavior of ${\cal I}_2$ in Eq.~(\ref{integrals}). 
In the limit $x/\tau \rightarrow 
\infty$ , the integration variable satisfies $\eta \gg1$ and thus we 
can use the asymptotic expression of the Green's function 
corresponding to the right, algebraic decaying tail, $G_{\alpha,1,-1,0} \sim 
\eta^{-(1+\alpha)}$ and get, after an integration by parts ,the 
asymptotic expansion
\bq
{\cal  I}_2 \sim \int_{x/\tau}^{\infty} e^{-\lambda \tau 
\eta} \, \eta^{-(1+\alpha)}\, d \eta =
\frac{\tau^\alpha}{\lambda}\frac{e^{-\lambda x}}{x^{1+\alpha}} - 
\frac{\alpha+1}{\lambda \tau} \int_{x/\tau}^{\infty}
e^{-\lambda \tau \eta} \, \eta^{-\left(\alpha+2\right)} \, d\eta \, .
\eq
Integrating by parts once more, gives the next term in the asymptotic 
series
\bq
\label{I2}
{\cal  I}_2 \sim \frac{\tau^\alpha}{\lambda}\frac{e^{-\lambda 
x}}{x^{1+\alpha}} \left[ 
1 - \frac{\left(\alpha+1\right)}{\lambda x}+ \ldots
\right] \, .
\eq
To deal with ${\cal  I}_1$, introduce a cut-off $\Omega$ such that $1 \ll \Omega$ and write
the integral as
\bq
\label{I1}
{\cal  I}_1={\cal C} + \int_{\Omega}^{x/\tau} \, 
e^{(\nu-\lambda) \tau \eta} \hat{G}_{\alpha,1,-1,0}(\eta) d \eta  \, ,
\eq
where the constant on the right hand side is defined as  ${\cal 
C}=\int_{-\infty}^{\Omega} \,  e^{(\nu-\lambda) \tau \eta} G_{\alpha,1,-1,0}(\eta) d \eta$.
Note that because of the faster than exponential decay of the 
asymmetric L\'evy distribution $G_{\alpha,1,-1,0}(\eta) $ for $\eta<0$ 
in Eq.(\ref{green_1}) the integrals converge for either sign of 
$\nu-\lambda$.
Since both the cut-off $\Omega$ and $x/\tau$ are assumed to be large, 
we can substitute as before the asymptotic expression of the Green's 
function in the integral of Eq.(\ref{I1}), and after an integration by parts 
obtain the asymptotic expansion
\bq
\label{I1_asy}
{\cal  I}_1 \sim {\cal C} + \frac{\tau^\alpha}{\nu-\lambda}\, 
\frac{e^{(\nu-\lambda)  x}}{x^{1+\alpha}}+ \dots  \, .
\eq
Substituting Eq.(\ref{I1_asy}) and Eq.(\ref{I2}) into Eq.(\ref{lead_edge})
we get 
\bq
\label{asym_lead_edge}
\phi \sim {\cal C} e^{-\nu x + \left(\gamma - \chi \lambda^\alpha 
\right) t}+ \left(\frac{1}{\nu-\lambda}+ \frac{A}{\lambda}\right)
\frac{\chi t}{x^{\alpha+1}} 
e^{-\lambda x + \left( \gamma -\chi 
\lambda^\alpha \right) t} \, .
\eq
for $\nu \neq \lambda$. 
The issue now is to determine the leading 
order term in Eq.~(\ref{asym_lead_edge}). The answer to this problem 
depends on the  relative values of 
$\nu$ and $\lambda$. 

If $\nu>\lambda>0$, i.e., if the initial condition decays faster than 
the truncation, the leading order term in 
Eq.~(\ref{asym_lead_edge}) for large $x$ is
\bq
\label{asym_tr}
\phi \sim \left[ \frac{1}{\nu - \lambda}+ 
\frac{A}{\lambda}\right ]\, 
\frac{ t \chi }{x^{1+\alpha}}\, e^{-\lambda x + \left( \gamma - \chi 
\lambda^\alpha \right) t} \, .
\eq
Note that, contrary to the $\alpha$-stable case, the asymptotic spatial decay of the front 
leading edge in Eq.~(\ref{asym_tr}) is independent of the value of $A$. 
The role of the truncation  is clearly seen in the exponential 
factor $e^{-\lambda x}$ that dominates the decay for $x \gg 
1/\lambda$. Figure~2-(c) shows a very good agreement between the 
numerical result and the scaling in Eq.~(\ref{asym_tr}).

When $x \ll 1/\lambda$ we can expand the exponential in 
Eq.~(\ref{asym_tr}) and write
\bq
\label{asym_tr_2}
\phi \sim \left[ \frac{1}{\nu - \lambda}+ 
\frac{A}{\lambda}\right ]\, t \chi 
\, e^{ \left( \gamma - \chi 
\lambda^\alpha \right) t} \, \frac{1}{x^{1+\alpha}}\, \left [ 
1-\lambda x + \frac{\lambda^2}{2} x^2 \ldots \right] \, .
\eq
According Eq.~(\ref{asym_tr_2}) in the intermediate asymptotic regime, $\left( \chi t \right)^{1/\alpha} \ll x 
\ll 1/\lambda$,  the front exhibits to leading order the ideal 
(untruncated) L\'evy flight algebraic scaling, $\phi \sim 
1/x^{1+\alpha}$. This scaling is numerically verified in Fig.~2-(b). 
Moreover, as Figs.~3, 4 and 5 show, in the intermediate asymptotic regime 
(which in the numerical simulations roughly corresponds to 
$0<\lambda \leq 1$) the front's velocity and acceleration  exhibit unbounded monotonic growth
and follow to a good approximation the ideal L\'evy flight scaling 
in Eqs.~(\ref{asy_x})-(\ref{asy_a}).

Going back to  Eq.~(\ref{asym_tr}) we have that outside the 
intermediate asymptotic regime, i.e. for $x>1/\lambda$,
the  Lagrangian trajectory of the front  is given by
\bq
\label{x_l}
-\lambda x_L(t) + \left( \gamma - \chi \lambda^\alpha\right) t + \ln t 
-(\alpha+1) \ln x_L(t) = M \, ,
\eq
where $M$ is a constant that depends on $\phi_0$. 
From Eq.~(\ref{x_l}) we obtain 
the following expression for the Lagrangian velocity of the front, 
$v_L=d x_L(t)/dt$,
\bq
\label{v_l}
v_L(t)=\frac{\gamma - \chi 
\lambda^\alpha+\frac{1}{t}}{\lambda + \frac{\alpha+1}{x_L(t)}}\, .
\eq
As shown in 
Figs.~\ref{fig_x_l} and  ~\ref{fig_v_l} there is  good agreement between the 
numerical results (solid lines) 
and Eqs.~(\ref{x_l}) and (\ref{v_l}) (dashed lines) in the asymptotic regime
$ x \gg \left( \chi t \right)^{1/\alpha}$.
From Eq.~(\ref{v_l}) it follows that in the limit $t \rightarrow 
\infty$ 
\bq
\label{v_l_decay}
\label{vv}
v_L \sim v_* -\frac{\alpha}{\lambda t}  \ldots \, ,
\eq
where the terminal velocity is given by
\bq
\label{v_l_term}
v_* = \frac{\gamma-\lambda^\alpha \chi}{\lambda}\, ,
\eq
which is positive since it has been assumed that $\gamma>\lambda^\alpha 
\chi$. 
The asymptotic approach to the terminal velocity is clearly observed in Fig.~\ref{fig_v_l} where 
the horizontal dashed lines show the terminal velocity in 
Eq.~(\ref{v_l_term}) for  the values of $\lambda$ 
considered in the numerical simulations. 
According to Eq.~(\ref{vv}) the time required for 
the front velocity to approach the terminal velocity within a given 
margin $v_*-v_L$ scales as $t \sim 1/\lambda$. 
The corresponding Lagrangian acceleration of the front, $a_L=d v_L(t)/dt$, is given by
\bq
\label{a_l}
a_L(t)=\frac{v_L(t)}{t\left( \lambda t v_*+1\right)}
\left\{ \left( \alpha+1\right) \left[\frac{v_L(t) 
}{x_L(t)/t}\right]^2-1\right\}
\, .
\eq
From Eqs.~(\ref{v_l}) and (\ref{a_l})   it follows that for large times,
\bq
\label{a_l_as}
a_L(t) \sim  \frac{\alpha}{\lambda t^2} \, .
\eq
As shown in Fig.~\ref{fig_v_l_a_l_sca}, the analytical scaling relations 
agree  well with the numerical results. 
In this figure, the curved dashed 
lines correspond to the analytical result in Eqs.~(\ref{v_l}) and 
(\ref{a_l}) and the  straight dashed lines correspond to the 
scaling in Eqs.~(\ref{v_l_term}) and (\ref{a_l_as}). 
Thus, outside the intermediate asymptotic regime, i.e. for $1/\lambda < x$, the front 
acceleration decays and the front 
approaches a constant terminal velocity as $t \rightarrow \infty$. 
However, the convergence of the dynamics to the constant front velocity 
regimen exhibits a very slow, $\sim 1/(\lambda t)$, algebraic decay 
compared to the significantly faster exponential convergence 
in the diffusive case. 

The calculations presented up to now assumed $\nu > \lambda$.
However, when $\lambda > \nu$, i.e., when the initial 
condition decays slower than the truncation, the leading order 
term in Eq.~(\ref{asym_lead_edge}) is
\bq
\label{asym_exp}
\phi \sim {\cal C} e^{-\nu x + \left(\gamma-\chi \lambda^\alpha 
\right) t}\, ,
\eq
which indicates that the front moves with the constant 
velocity 
\bq
\label{front_speed}
v= \frac{\gamma-\lambda^\alpha \chi}{\nu}\, .
\eq
An example of this  over-truncated regime is presented in 
Fig.~\ref{fig_fronts}-(d) that  shows a very good agreement with the 
exponential decay in 
Eq.~(\ref{asym_exp}) for $\lambda=100$ and $\nu=50$.
The remaining case to consider is $\lambda=\nu$. In this case, the asymptotic approximation in 
Eq.~(\ref{I2}) still holds. However, the expression in Eq.~(\ref{I1_asy}) can not be 
used, and we have to go back to the integral ${\cal I}_1$ Eq.~(\ref{I1})
\bq
\label{I_1_b}
{\cal  I}_1={\cal C} + \int_{\Omega}^{x/\tau} \, 
G_{\alpha,1,-1,0}(\eta) d \eta  \, ,
\eq
where this time the constant on the right hand side is defined as  ${\cal 
C}=\int_{-\infty}^{\Omega} \,   \hat{G}_{\alpha,1,-1,0}(\eta) d \eta$.
Using the asymptotic expression for $\hat{G}_{\alpha,1,-1,0} \sim 
\eta^{-1-\alpha}$ and integrating, it is concluded that the leading order 
term in  Eq.~(\ref{lead_edge}) is
$\phi \sim {\cal C} e^{-\lambda x + \left(\gamma-\chi \lambda^\alpha 
\right) t}$ which implies that the constant velocity of the front is 
given by Eq.(\ref{front_speed}) with $\nu=\lambda$.

\section{Conclusions}

We have presented a numerical and 
analytical study of the role of truncated L\'evy flights in the 
propagation of fronts in reaction superdiffusion systems. The study 
was based on the truncated fractional Fisher-Kolmogorov model in 
which  the spatial derivative  is replaced by the 
$\lambda$-truncated fractional derivative of order $\alpha$. 

Depending on the level of truncation four 
front propagation regimes can be distinguished: an asymptotic algebraic 
regime, an intermediate asymptotic algebraic 
regime, a truncated regime, and an over-truncated 
regime. The asymptotic algebraic regime corresponds to $\lambda=0$. In this 
case the problem reduces to the $\alpha$-stable fractional 
Fisher-Kolmogorov problem that exhibits exponential acceleration and 
algebraic decaying tails for $x \gg \left( \chi t \right)^{1/\alpha}$.
The intermediate asymptotic regime is characterized by $\left( \chi t \right)^{1/\alpha} \ll x 
\ll 1/\lambda$. We have shown numerically and analytically 
that in this regime the truncation effects are negligible and 
the algebraic decay of the tail as well as the acceleration of the front prevail. 
Outside the intermediate asymptotic regime, i.e. for $x > 1/\lambda$ 
and $\left( \chi t \right)^{1/\alpha} \ll x$, the  tail of the front  exhibits the tempered decay 
$\phi \sim   e^{-\lambda x}/ x^{1+\alpha}\,$, the acceleration is 
transient, and the front eventually reaches a terminal speed as $t\rightarrow \infty$. 
In the over-truncated regime the  truncation decays  faster 
than the initial condition, i.e. $\lambda> \nu$. In this case,  L\'evy flights 
have apparently no qualitative effect on the asymptotic dynamics of the front 
that exhibits a diffusive-type exponential tail, $\phi \sim e^{-\nu 
x}$, and constant propagation velocity, $v=\gamma_{eff}/\nu$. 
However, contrary to the diffusive case, the constant velocity in the 
over-truncated case is a monotonically decaying function of $\nu$ 
and has no finite minimum. 

Although in the truncated regime the acceleration decays and the front 
eventually reaches a constant terminal speed, the numerical and 
analytical results show that the convergence of this process is very 
slow. In particular the front acceleration asymptotically decays as
$a_L \sim \alpha  /\left( \lambda t^{2}\right)$  and 
the approach to the terminal velocity scales as $v_L \sim v_* - \alpha 
/\left( \lambda t \right)$.  This algebraic convergence is in sharp 
contrast with the exponential convergence observed in the diffusive
 Fisher-Kolmogorov equation. In this sense 
the  truncated regime  resembles the ``ultra-slow'' convergence 
regimen  observed in transport problems without reaction terms.  

One of the motivations and potential applications of the present work lays on the study of 
transport in magnetically confined plasmas. In this system it 
 has been suggested that reaction diffusion models provide insightful, 
 though highly simplified,  reduced descriptions of the interaction of 
 turbulence and transport.  Current models usually 
 assume Laplacian diffusive operators, see for example Refs.~\cite{rd_plasmas}. However
 there is evidence that transport in magnetically confined plasmas is 
 not necessarily diffusive, see for example \cite{iter} and 
 references therein. On the other hand, it has been argued that 
 truncated L\'evy distributions might play a role in the description 
 of electrostatic potential fluctuations \cite{kaw}. 
Based on this, it would of interest to explore the implications of the 
 present work on the corresponding plasma transport models.
 
Throughout this paper we have limited attention to extremal,
$\theta=-1$, transport processes and assumed an external advection 
velocity to cancel the drift resulting from the truncation of the 
asymmetric fractional derivative. A problem of interest is to 
perform more general numerical simulations considering different 
degrees of asymmetry and including memory effects
through the use of fractional derivatives in time.
As mentioned in the introduction in recent years several papers have 
discussed the role of Le\'evy flights in front propagation and 
pattern formation in reaction-anomalous-diffusion systems. It would be of interest to explore 
the role of truncation effects in these systems.
Beyond its intrinsic theoretical interest, the study presented here 
might have relevance in the interpretation and modeling of laboratory 
experiments and numerical simulations of complex systems. It is 
plausible that in these systems,  
the presence of boundary conditions, finite size 
domain and decorrelation could introduce ``non-ideal'' L\'evy-flights 
dynamics of the type discussed here.

\subsection*{Acknowledgments}

This work has been supported by  the Oak Ridge
National Laboratory, managed by UT-Battelle, LLC, for the U.S.
Department of Energy under contract DE-AC05-00OR22725.

%

\pagebreak
\newcounter{figlist}
\subsection*{FIGURE CAPTIONS}
\begin{list}
{FIG.~\arabic{figlist}.}{\usecounter{figlist}
                     \setlength{\labelwidth}{.55in}
                     \setlength{\leftmargin}{.55in}}


\item
\label{fig_greens}
(Color online) 
Green's functions of the asymmetric truncated fractional diffusion 
Eq.~(\ref{trunc_green}) for  $\alpha=1.5$, $\theta=-1$, and 
$\lambda=0$, $10$, and 
$20$. The solid (blue) lines show $G$ as function of $x$ for fixed $t$, and 
the dashed (black) lines show the corresponding asymptotic dependence 
according to Eq.~(\ref{trunc_green_as}). The numerical and 
asymptotic results are practically indistinguishable for $x>0.1$ and 
for visualization purposes the asymptotic result has been shifted 
downward a little bit in the plot. 

\item
\label{fig_fronts}
(Color online) 
Depending on the value of the truncation parameter $\lambda$ four 
front propagation regimes can be distinguished in the truncated fractional 
Fisher-Kolmogorov Eq.~(\ref{tffk}): (a) Asymptotic algebraic 
regime for $\lambda=0$; (b) Intermediate asymptotic  algebraic 
regime for $\lambda \neq 0$; (c) Truncated regime for $0<\lambda<\nu$; (d) Over-truncated 
regime for $\lambda>\nu$. 
In all four panels the left most (red) curve 
denotes the initial condition in Eq.~(\ref{ic}) and the other (blue) solid line curves show the front 
profiles at different  times. The dashed (red) curve shows the 
analytical asymptotic result according to Eq.~(\ref{alpha-stable}) 
with $A=0$ in panels (a) and (b); according to Eq.(\ref{asym_tr}) in 
panel (c); and according to Eq.~(\ref{asym_exp}) in panel (d). 
In all cases $\alpha=1.5$, $\theta=-1$, $\gamma=1$, and 
$\chi=5\times 10^{-7}$.

\item
\label{fig_x_l}
(Color online)
Space-time  Lagrangian front paths, $\phi(x_L(t),t)=\phi_0$ with 
$\phi_0=10^{-6}$, according to the
asymmetric truncated fractional Fisher-Kolmogorov 
Eq.~(\ref{tffk}) with  $\alpha=1.5$, $\theta=-1$, $\gamma=1$, 
$\chi=5\times 10^{-7}$ and different 
values of $\lambda$. The solid (blue) curves denote the numerical 
results and the dashed (red) curves the asymptotic result according 
to Eq.~(\ref{x_l}). The dotted (green) lines denote the front speed 
(upper line) and the minimum front speed (lower line) in
the Gaussian diffusive case according to Eq.~(\ref{diff_speed}). 
As the $\lambda=1$ case shows,  the effect of 
truncation is negligible in the intermediate asymptotic regime. 

\item
\label{fig_v_l}
(Color online)
Time dependence of  Lagrangian front velocities, $v_L(t)=d x_L(t)/dt$, according to the
asymmetric truncated fractional Fisher-Kolmogorov 
Eq.~(\ref{tffk}) with  $\alpha=1.5$, $\theta=-1$, $\gamma=1$, 
$\chi=5\times 10^{-7}$ and different 
values of $\lambda$. The solid (blue) curves denote the numerical 
results and the dashed (red) curves the asymptotic result according 
to Eq.~(\ref{v_l}). The horizontal (black) dash lines denote the 
corresponding terminal velocities according to Eq.~(\ref{v_l_term}).
The solid (green) line at the bottom denotes the corresponding 
front speed  in the Gaussian diffusive case according to Eq.~(\ref{diff_speed}). 

\item
\label{fig_a_l}
(Color online)
Dependence of front acceleration on truncation. The panels show the
time dependence of  the Lagrangian front acceleration $a_L(t)=d v_L(t)/dt$, according to the
asymmetric truncated fractional Fisher-Kolmogorov 
Eq.~(\ref{tffk}) with  $\alpha=1.5$, $\theta=-1$, $\gamma=1$, 
$\chi=5\times 10^{-7}$ and different 
values of $\lambda$. The solid (blue) curves denote the numerical 
results and the dashed (red) curves the asymptotic result according 
to Eqs.~(\ref{asy_a})  and (\ref{a_l}). In the intermediate asymptotic regime, panels 
(a)-(c), the front exhibit monotonically increasing acceleration. In the truncated 
regime, panels (d)-(f) the acceleration is transient and decays at 
large times. 

\item
\label{fig_v_l_a_l_sca}
(Color online)
Asymptotic algebraic scaling of approach to terminal velocity, 
$v_*-v_L$, and asymptotic algebraic scaling of Lagrangian front acceleration 
decay . The solid (blue) curves denote the numerical 
results for $\alpha=1.5$, $\theta=-1$, $\gamma=1$, 
$\chi=5\times 10^{-7}$ and different 
values of $\lambda$. The dashed (red) curves show the asymptotic results according 
to Eqs.~(\ref{v_l}) and (\ref{a_l}), and the dashed (black) straight 
lines the analytical asymptotic scalings, $v_*-v_L \sim 
\alpha/(\lambda t)$ and $a_L \sim \alpha/(\lambda t^2)$ according to 
Eqs.~(\ref{v_l_decay}) and (\ref{a_l}). 

\end{list}

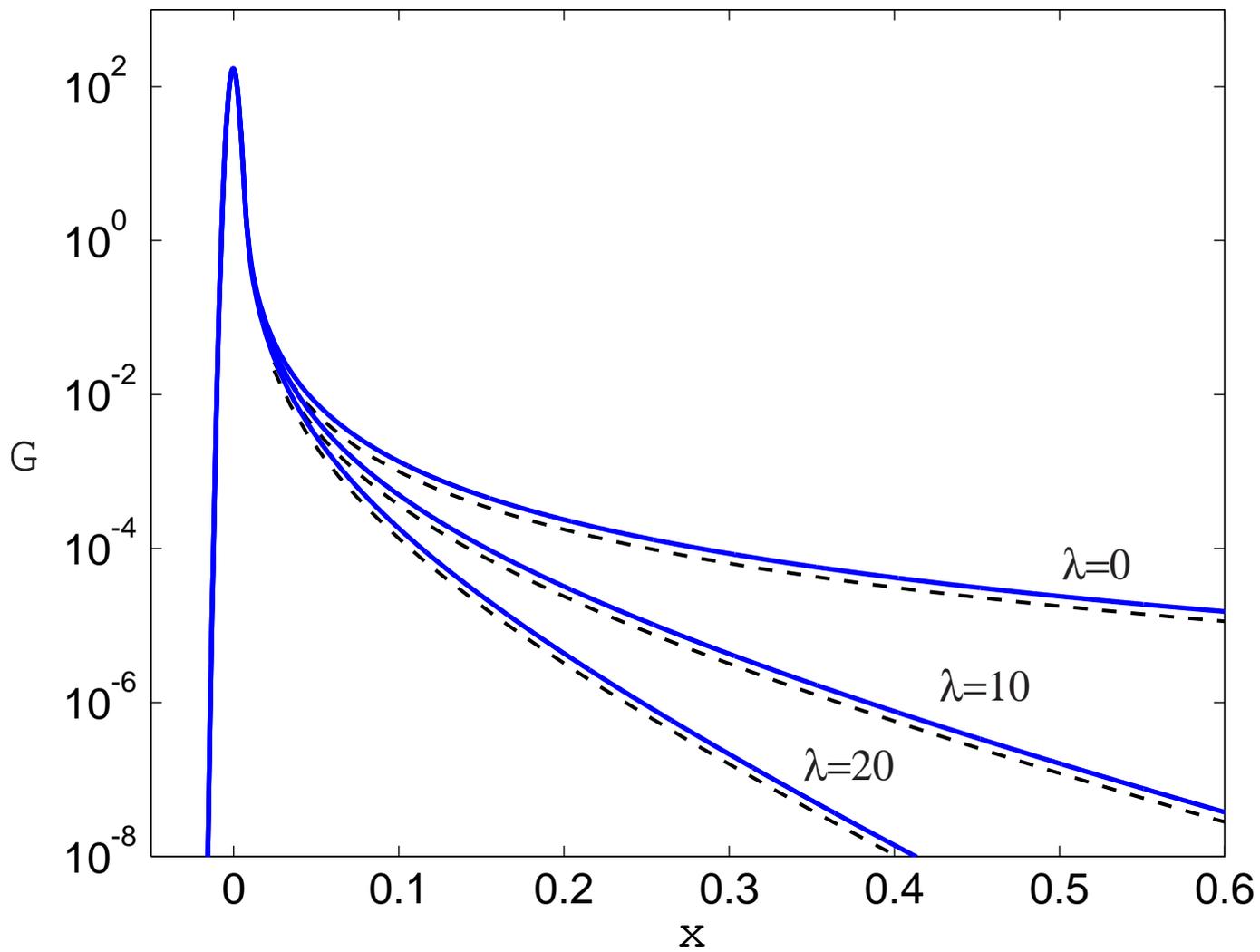
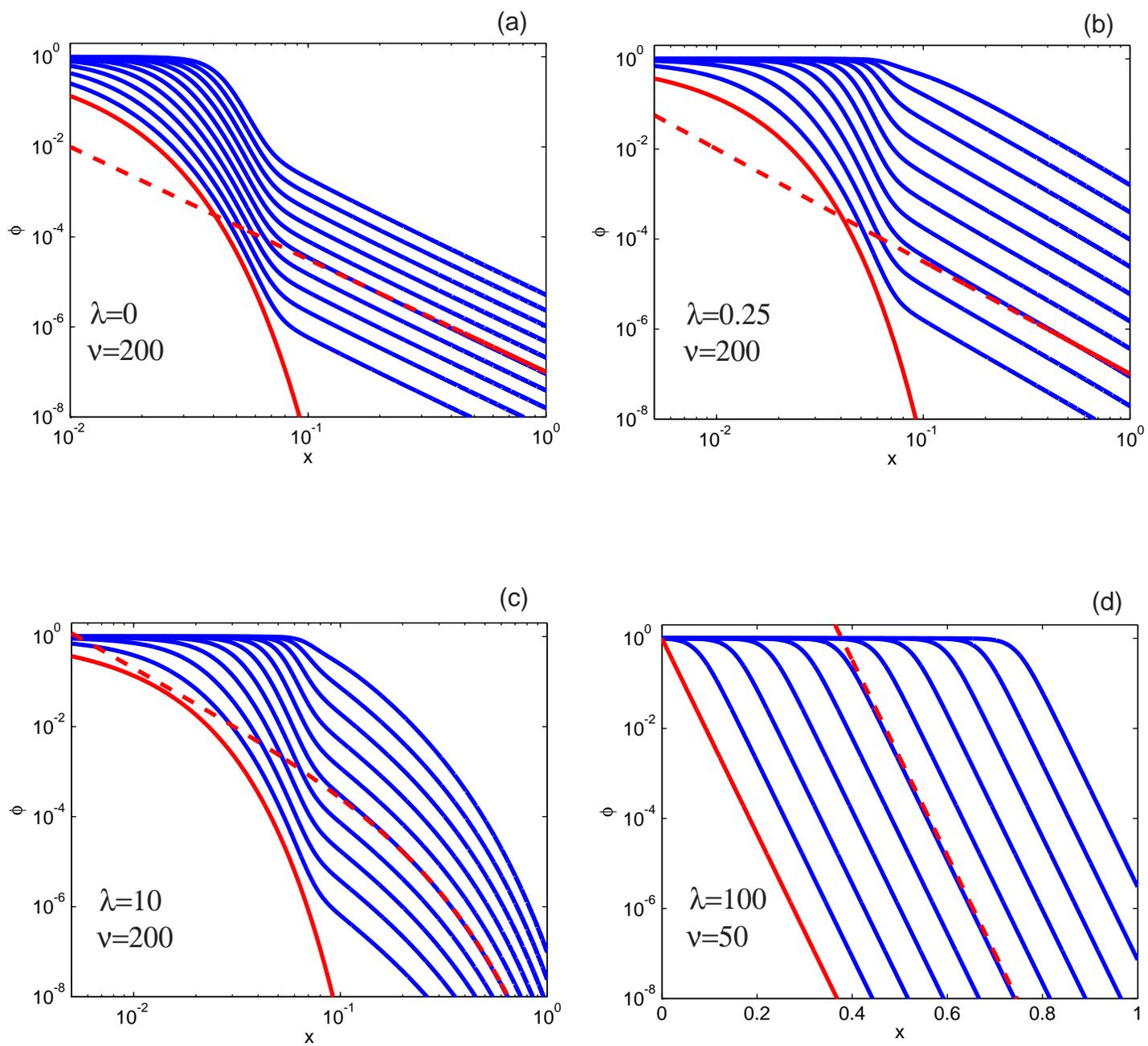
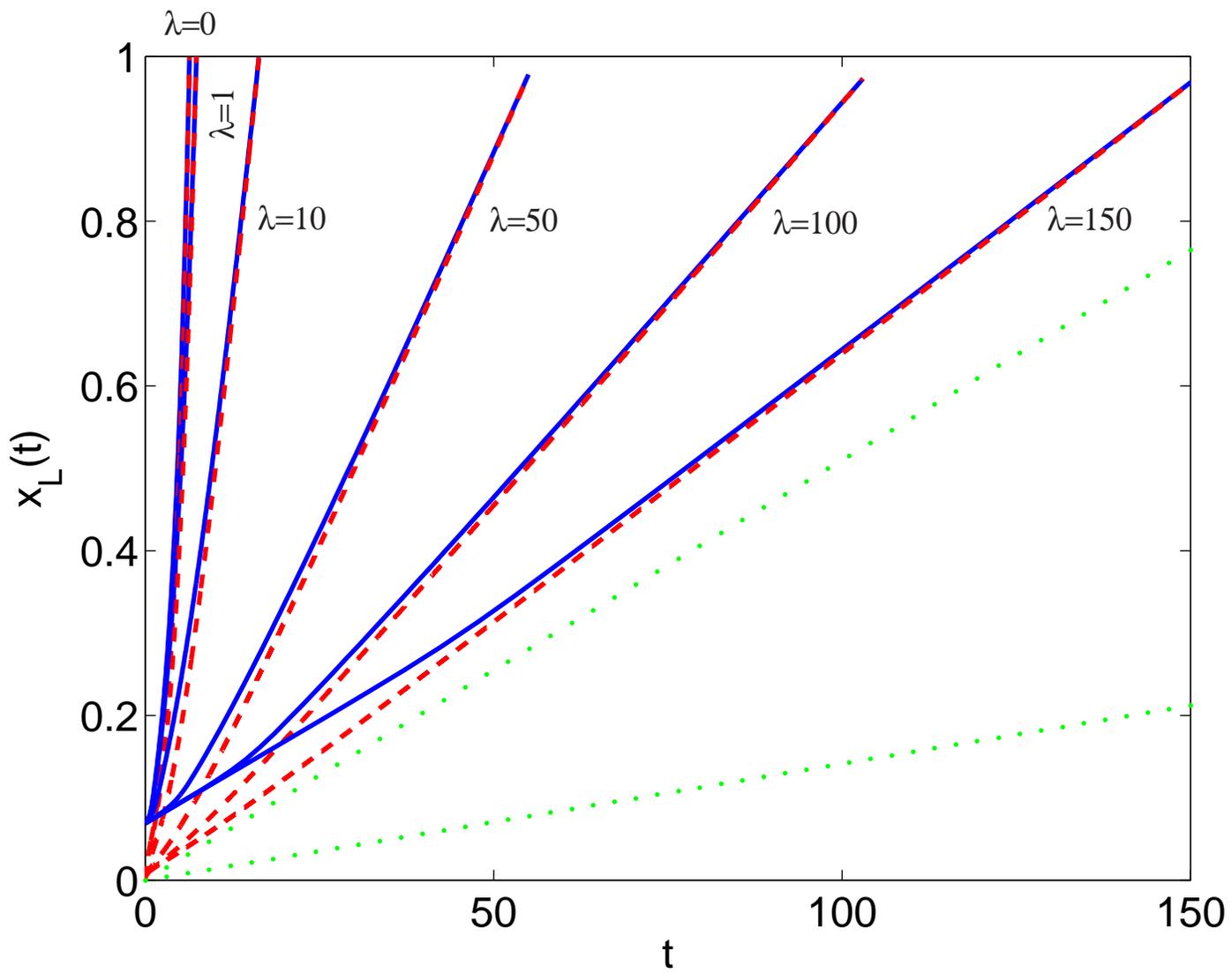
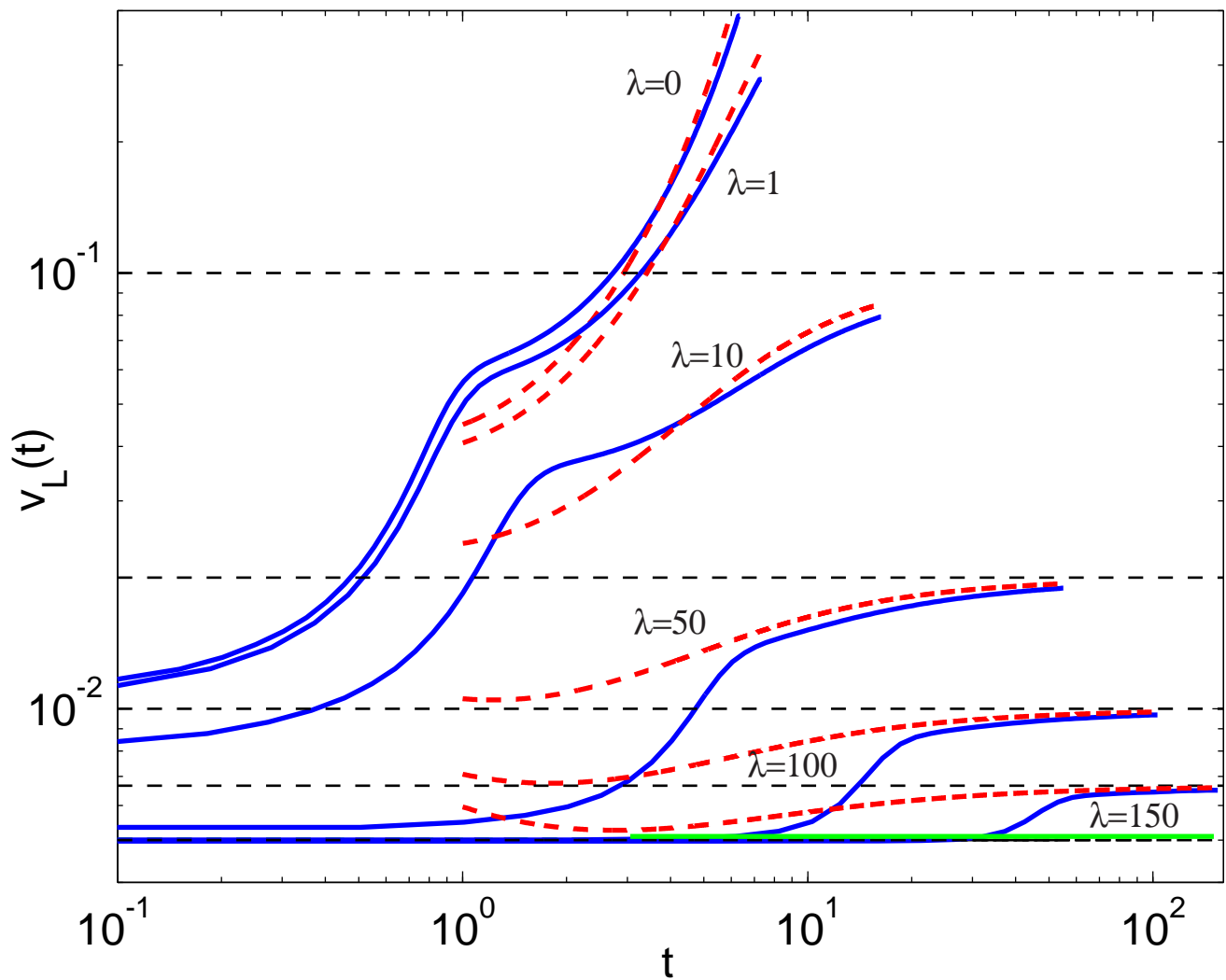
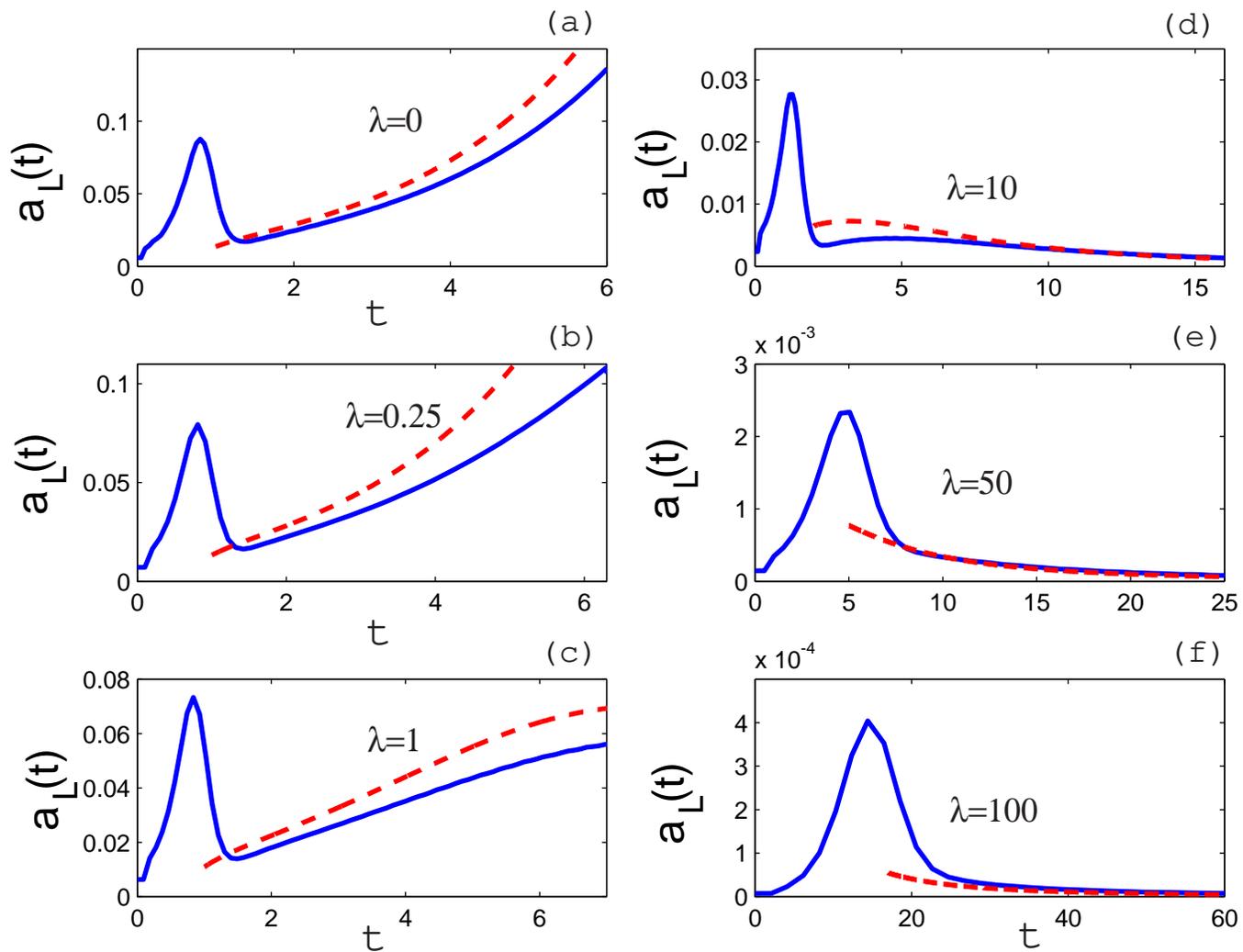
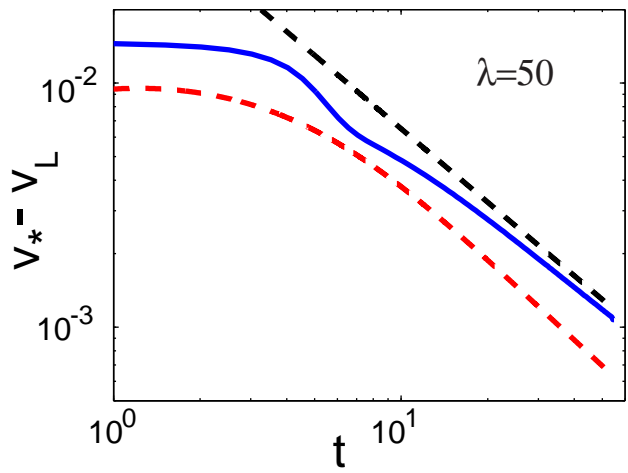
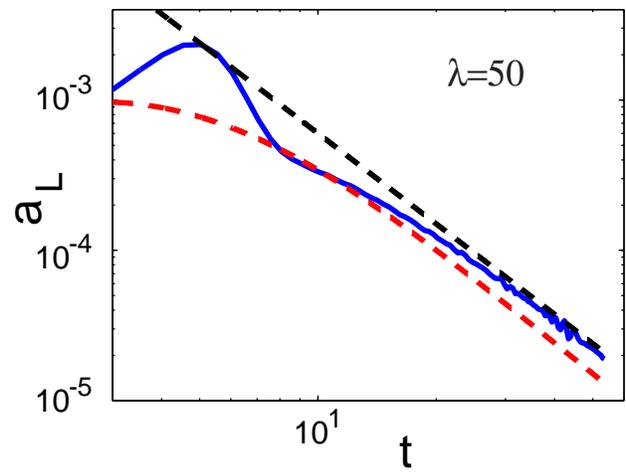
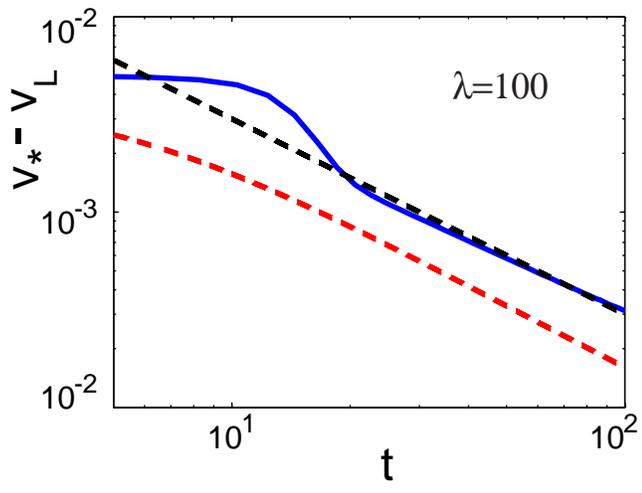
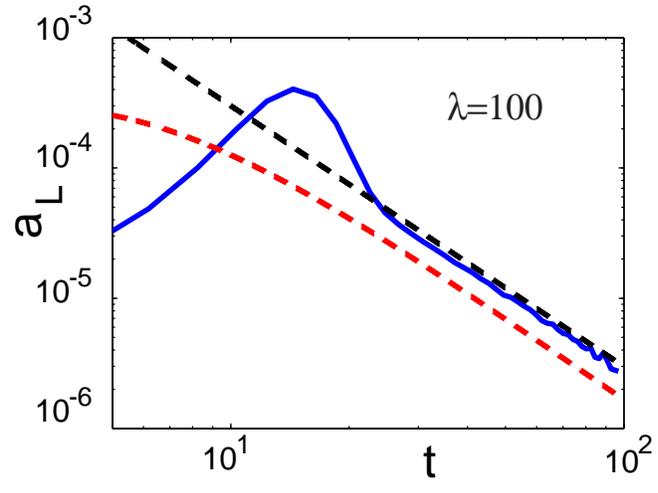
\end{document}